%% file: main.tex
\documentclass[conference]{IEEEtran}
\ifCLASSINFOpdf
\else
\fi
\usepackage{url}

\usepackage{xcolor}
\usepackage{graphicx}
\usepackage{enumitem}
\usepackage{comment}

\usepackage[most]{tcolorbox}
\tcbset{
    colback=gray!5,
    colframe=gray!75,
    boxrule=0.5pt,
    arc=3pt,
    left=6pt,
    right=6pt,
    top=6pt,
    bottom=6pt
}

\begin{document}
%
\title{CAN-QA: A Question-Answering Benchmark for Reasoning over In-Vehicle CAN Traffic}

\author{
  Jing Chen$^{*}$,
  Abhijay Deevi$^{*}$, 
  Onat Gungor,
  Tajana Rosing\\
  \textit{Department of Computer Science and Engineering} \\
  \textit{University of California, San Diego (UCSD)} \\
  \{jic128, adeevi, ogungor, tajana\}@ucsd.edu\\
}

\maketitle
\begingroup\renewcommand\thefootnote{$*$}\footnotetext{Both authors contributed equally to this research.}\endgroup


%


\newcommand{\Design}[0]{\textsc{CAN-QA}}

\maketitle

\begin{abstract}
\input{Abstract}
\end{abstract}


%
\IEEEpeerreviewmaketitle

\section{Introduction}
\input{Introduction}

\section{Related Work}
\input{RelatedWork}

\section{\Design{} Framework}
\input{Framework}

\section{Results}
\input{Results}

\section{Conclusion}
\input{Conclusion}

\section*{Acknowledgements}
This work has been funded in part by NSF, with award numbers \#2112665, \#2112167, \#2003279, \#2120019, \#2211386, \#2052809, \#1911095 and in part by PRISM and CoCoSys, centers in JUMP 2.0, an SRC program sponsored by DARPA.



\bibliographystyle{IEEEtran}
%


\bibliography{biblio}

\end{document}

%% file: Abstract.tex
The Controller Area Network (CAN) is a safety-critical in-vehicle communication protocol that lacks built-in security mechanisms, making intrusion detection essential. Existing approaches predominantly formulate CAN intrusion detection as a classification task, mapping complex traffic patterns to attack labels. However, this formulation abstracts away the temporal and relational structure of CAN traffic and misaligns with real-world forensic workflows, which require systematic reasoning about traffic behavior. To address this gap, we introduce \Design{}, the first benchmark that reformulates CAN traffic analysis as a question-answering (QA) task. \Design{} converts raw CAN logs into temporally segmented windows and applies deterministic rule-based templates to generate natural-language questions paired with automatically derived ground-truth answers. The resulting dataset comprises 33,128 QA pairs across 10 categories, each targeting distinct semantic and temporal properties of CAN traffic. Using \Design{}, we evaluate large language models across both True/False and multiple-choice formats. Our results indicate that, although these models capture superficial statistical regularities, they struggle with temporal reasoning, multi-condition inference, and higher-level behavioral interpretation. Our code is available at \url{https://github.com/Kriiiiss/CAN-QA}. 

%% file: Introduction.tex
Modern vehicles rely on in-vehicle communication networks to coordinate safety-critical components, including braking, steering, engine control, and advanced driver assistance systems. Among these networks, the Controller Area Network (CAN) remains the dominant in-vehicle communication protocol due to its simplicity, efficiency, and low deployment cost \cite{zeng2016vehicle}. However, CAN was originally designed without security considerations~\cite{tang2025wip}. It provides no authentication or message integrity protection and operates as a broadcast medium in which any compromised electronic control unit (ECU) can transmit messages onto the bus~\cite{lotto2024survey}. These fundamental limitations expose CAN networks to a wide range of cyberattacks, making intrusion detection a central and persistent challenge in automotive cybersecurity~\cite{serag2025sok,lampe2023intrusion}.

\begin{figure}[]
    \centering
    \includegraphics[width=.49\textwidth]{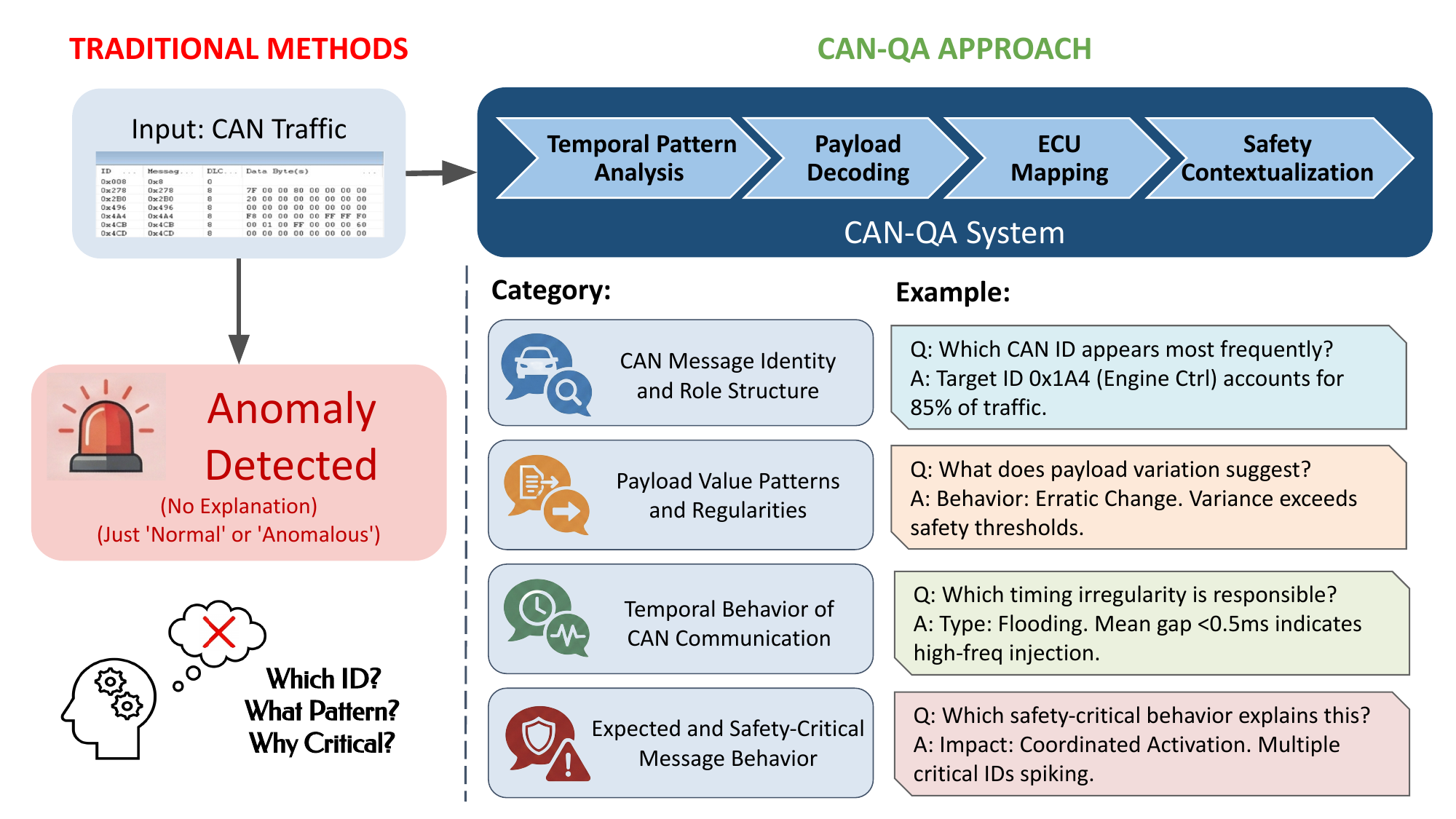}
    \caption{Advancing Beyond Binary Anomaly Detection to \Design{}.}
    \label{fig:Motivation}
\end{figure}


Most existing intrusion detection approaches for automotive CAN networks formulate the problem as statistical anomaly detection or supervised classification over raw CAN signals \cite{tanksale2024intrusion,hoang2024supervised,lokman2019intrusion}. While effective at flagging deviations from normal behavior, these approaches provide limited insight into how and why anomalies occur, reducing interpretability and forensic utility \cite{jeong2023x,neupane2022explainable}. In practice, CAN traffic analysis is inherently investigative and question-driven: analysts aim to localize anomalous behavior, examine timing and payload irregularities, and relate observed patterns to known attack strategies rather than relying solely on binary detection~\cite{ozdemir2024survey}.

Recent advances in LLMs have demonstrated strong capabilities in structured reasoning and natural-language interpretation, enabling them to support analyst-driven workflows \cite{wei2022chain}. However, existing benchmarks for automotive intrusion detection focus on detection accuracy and dataset coverage, and do not evaluate whether models can reason about CAN traffic patterns \cite{lee2023comprehensive}. This limitation motivates a new problem formulation that moves beyond binary detection and instead frames CAN intrusion analysis as a structured question-answering (QA) task, with explicit reasoning targets that are interpretable and aligned with analyst needs. Figure~\ref{fig:Motivation} contrasts traditional anomaly-only detection with our question-driven formulation, illustrating how CAN-QA transforms raw CAN traffic into structured analytical queries such as identifying dominant IDs, diagnosing timing irregularities, interpreting payload behavior, and contextualizing safety impact, replacing opaque binary alerts with explicit, multi-dimensional reasoning outputs.

In this work, we introduce \Design{}, the first QA framework and benchmark that reformulates CAN traffic analysis as a structured reasoning problem. \Design{} converts raw CAN logs into fixed-length temporal windows and applies deterministic, rule-based templates to generate natural-language questions with automatically derived ground-truth labels. Each question corresponds to a window-level property of CAN communication that can be computed directly from the observed frames, ensuring consistency, reproducibility, and eliminating the need for manual annotation. The framework is dataset-agnostic: by recomputing baseline statistics from new CAN traces, it can generate QA datasets for different vehicles, attack scenarios, or data sources. Using this approach, we instantiate \Design{} as a benchmark that standardizes reasoning-based evaluation over CAN traffic, shifting the focus from binary detection to interpretable, question-driven analysis.

Using \Design{}, we benchmark various LLMs on their ability to reason over CAN traffic. Our evaluation goes beyond detection accuracy, measuring model performance across diverse analytical dimensions, including temporal patterns, payload dynamics, and safety-critical message activity. We find that, while models capture surface-level statistical cues effectively, they consistently struggle with temporal reasoning, multi-condition inference, and higher-level behavioral interpretation. These findings reveal fundamental limitations of current LLMs in safety-critical cyber-physical systems and underscore the need for future research that integrates structured time-series reasoning with language-based analysis.

%% file: RelatedWork.tex
Intrusion detection for the in-vehicle CAN has received significant attention as modern vehicles become increasingly software-defined and interconnected. Early research primarily relied on statistical anomaly detection techniques that model message entropy, frequency, or timing characteristics of CAN traffic~\cite{muter2011entropy}. More recent work has shifted toward deep learning-based approaches that capture complex temporal dependencies and contextual relationships in CAN communication~\cite{tanksale2024intrusion}. Semi-supervised and adversarial strategies have also been proposed to handle the limited availability of labeled attack data. For example, the convolutional adversarial autoencoder~\cite{lokman2022canids} models CAN frame distributions to detect anomalous traffic patterns under partial supervision, while transformer-based methods such as CAN-BERT~\cite{kang2022canbert} leverage masked language modeling to learn effective representations of CAN payload sequences without handcrafted features.

The development of robust CAN intrusion detection methods has been supported by publicly available datasets, which provide standardized benchmarks for model evaluation. The ROAD dataset~\cite{verma2020road} offers a comprehensive real-world benchmark collected in a controlled dynamometer environment, and the Car Hacking dataset~\cite{kang2016carhacking} provides labeled normal and malicious CAN traffic. While these datasets have advanced supervised and semi-supervised detection, they focus on classification and anomaly detection and do not enable evaluation of higher-level reasoning or natural-language understanding of CAN traffic. \Design{} addresses this gap by introducing a QA-driven framework designed to evaluate reasoning over automotive communication patterns and attack signatures.

Beyond automotive security, recent research has explored large language models (LLMs) for structured reasoning over time-series data. Such sequential reasoning is directly applicable to CAN traffic, where messages exhibit temporal dependencies and multi-step relationships. Benchmarks such as MTBench~\cite{chen2025mtbench}, Time-MQA~\cite{kong2025time}, ITFormer~\cite{wang2025itformer}, and ChatTS~\cite{xie2024chatts} demonstrate that question-answering (QA) formulations provide a powerful interface for evaluating temporal understanding, contextual inference, and multi-step reasoning. By converting numerical sequences into natural-language reasoning tasks, these approaches support evaluation paradigms that go beyond conventional anomaly detection metrics.

Despite these advances, existing time series QA frameworks fail to capture key CAN specific characteristics, including bounded 8 byte payloads, strict timing dependencies, and safety critical operational constraints, which pose unique challenges for reasoning over vehicular data. Moreover, current automotive IDS research remains primarily focused on anomaly detection and classification, lacking mechanisms for evaluating reasoning, interpretability, or natural-language explanations of CAN behavior. \Design{} bridges this gap by providing a QA-driven benchmark for CAN intrusion detection, enabling systematic evaluation of whether modern language models can reason about communication patterns and interpret attack signatures through structured natural-language queries.

%% file: Framework.tex
\begin{figure}[!t]
    \centering
    \includegraphics[width=.49\textwidth]{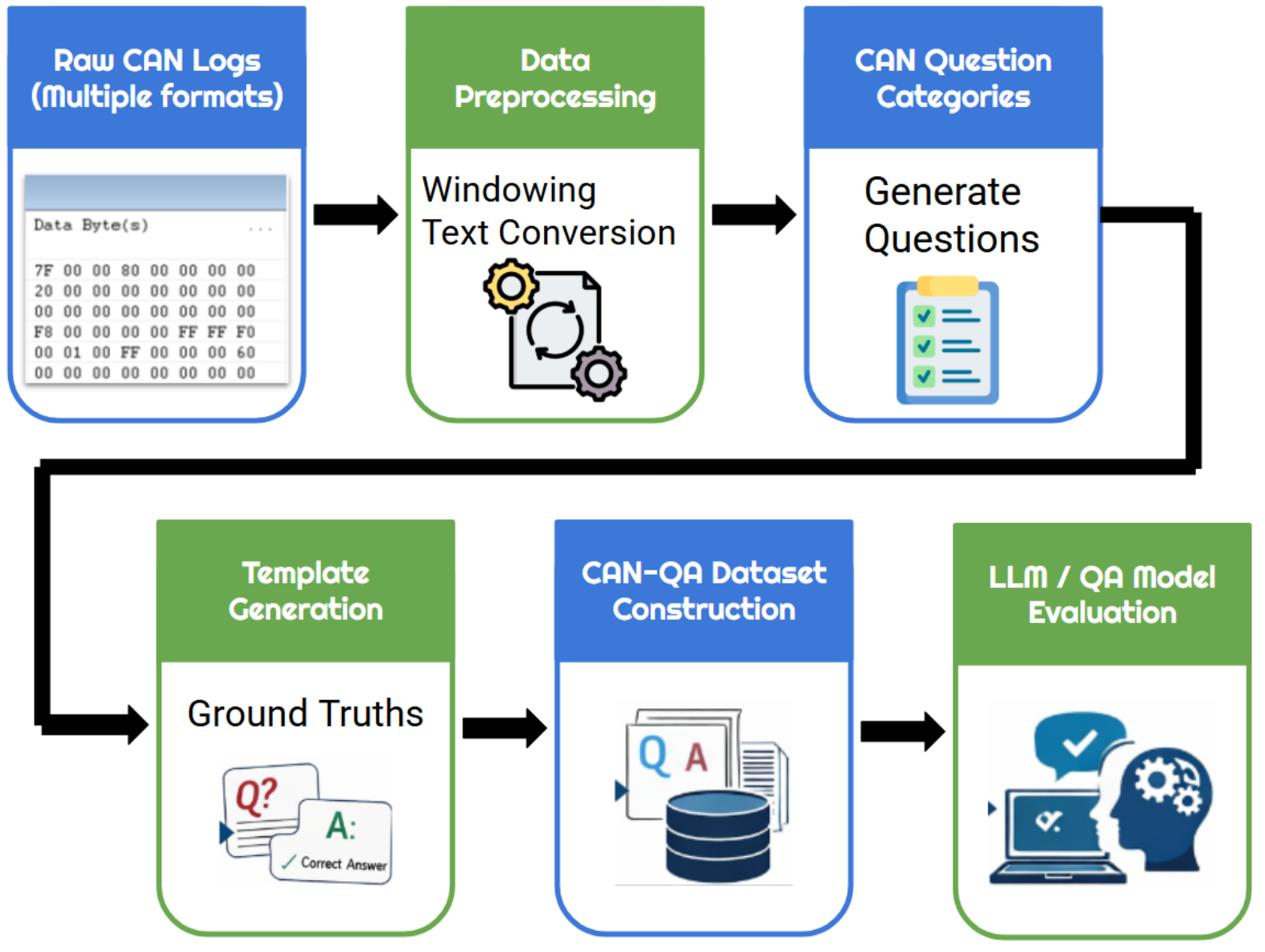}
    \caption{Overview of the \Design{} framework.}
    \label{fig:framework}
\end{figure}

Figure~\ref{fig:framework} illustrates the overall architecture of the \Design{} framework, which is designed to automatically generate cybersecurity-focused question–answer pairs from Controller Area Network (CAN) traffic. The framework first preprocesses raw CAN logs by normalizing heterogeneous data sources into a unified frame representation and organizing frames into temporally consistent windows. Each window is then transformed into a structured textual context that preserves both semantic and temporal properties of CAN communications. Using this contextual representation, the QA construction module generates diverse reasoning tasks, including traffic interpretation, anomaly detection, and temporal analysis questions. Finally, the generated question–answer pairs derived from CAN traffic are validated and structured to produce a high-quality dataset suitable for training and evaluating language models.

\subsection{Data Preprocessing}
We first normalize raw CAN logs into a unified frame schema consisting of a timestamp, identifier, data length code (DLC), eight payload bytes, and a standardized flag field. This step converts heterogeneous log formats into consistent field types, removes incomplete or malformed entries, and orders all frames chronologically to form a temporal sequence. 

The normalized CAN stream is then partitioned into fixed-length windows, where each window contains a contiguous sequence of CAN frames. Windowing provides a controlled temporal context for question--answer generation by constraining reasoning tasks to localized segments of bus activity. A fixed window size is adopted to maintain consistency across datasets and question categories. The window length is chosen to balance two competing objectives: capturing meaningful short-term communication dynamics while keeping the resulting textual context tractable for language models.

Each window is converted into a structured textual representation. Specifically, each CAN frame is formatted as a labeled line containing its timestamp, identifier, DLC, payload bytes, and flag field. These lines are concatenated in chronological order to form the final context presented to the QA model. This representation preserves the semantic content and temporal ordering of the original CAN trace while providing a human-readable, model-agnostic interface that enables language-based reasoning over raw communication activity.


\subsection{\Design{} Question Categories}
The question categories define a structured mapping from preprocessed CAN windows to executable question templates by organizing window-level statistics and baseline-relative signals into a fixed set of analyzable reasoning dimensions. The categories are designed to partition CAN communication properties into orthogonal behavioral axes, ensuring that each category captures a distinct and deterministically computable aspect of CAN activity. This design constrains question generation while guaranteeing coverage of observable window-level behaviors and ensuring that every instantiated question admits a unique and reproducible ground-truth answer.

The categories are constructed by grouping signals according to the CAN properties they characterize, including message participation, traffic distribution, temporal behavior, payload structure, protocol-level integrity, and multi-signal behavioral interpretation. Collectively, these categories span complementary dimensions of CAN communication while minimizing redundancy between question types. Based on this construction, we define ten CAN question categories employed throughout the framework, each accompanied by a representative example query. 


\textbf{Category 1: Anomalous Behavior Identification and Localization.}
This category focuses on determining the presence of anomalous frames within a Controller Area Network (CAN) window and characterizing their temporal distribution. It provides both binary anomaly detection and fine-grained localization signals that serve as grounding evidence for downstream reasoning processes. By establishing whether and where abnormal behavior occurs, this category forms the primary contextual foundation for higher-level analytical tasks.

\begin{tcolorbox}

\textbf{Question:} 
Given a window where one randomly chosen frame's Flag is hidden, is the hidden frame labeled as an attack with a non-zero Flag?

\textbf{Ground Truth:} 
\textit{True}

\end{tcolorbox}


\textbf{Category 2: CAN Message Identity and Role Structure.}
Category 2 characterizes which CAN identifiers are active within a window and evaluates deviations from expected identity participation patterns. It captures structural changes in message presence and functional role composition that cannot be inferred solely from traffic volume or timing behavior.

\begin{tcolorbox}

\textbf{Question:} 
What best explains the presence of many single-appearance CAN IDs in this window?

\textbf{Options:}
\begin{enumerate}[label=\Alph*.]
\item Possible fuzzing or probing behavior (the single-appearance share exceeds 30\% and unexpected ID count is greater than 0).
\item Normal background variation (the single-appearance share is at most 30\%).
\item Logging truncation or window boundary effects (the single-appearance share exceeds 30\% and unexpected ID count is 0 and the max ID share is below 50\%).
\item A single dominant ID masking others (the single-appearance share exceeds 30\% and the max ID share is at least 50\%).
\end{enumerate}

\textbf{Ground Truth:} 
\textit{B}
\end{tcolorbox}


\textbf{Category 3: Traffic Volume and Distribution Characteristics.}
Category 3 analyzes how message traffic is distributed across CAN identifiers within a window. It identifies dominance, imbalance, and concentration patterns that arise when communication load disproportionately shifts toward specific message identifiers.

\begin{tcolorbox}

\textbf{Question:} 
Is the window's frame rate (frames per second) above the 95th percentile of the dataset baseline?

\textbf{Ground Truth:} 
\textit{False}

\end{tcolorbox}


\textbf{Category 4: Temporal Behavior of CAN Communication.}
This category analyzes the temporal structure of CAN traffic by examining frame ordering, inter-arrival timing, and burst dynamics. It identifies timing-based anomalies that occur independently of message frequency or payload content, thereby enabling evaluation of temporal reasoning over sequential communication patterns.
\begin{tcolorbox}
\textbf{Question:} 
Which timing pattern is most consistent with this window?

\textbf{Options:}
\begin{enumerate}[label=\Alph*.]
\item Suppression-like behavior (at least one gap exceeds 0.001 seconds).
\item Flooding-like behavior (all gaps are below 0.0005 seconds).
\item Mixed or ambiguous timing signals (at least one gap exceeds 0.001 seconds and at least one gap is below 0.0005 seconds).
\item Normal periodic traffic (no gap exceeds 0.001 seconds and not all gaps are below 0.0005 seconds).
\end{enumerate}

\textbf{Ground Truth:} 
\textit{C}
\end{tcolorbox}


\textbf{Category 5: Payload Value Patterns and Regularities.}
Category 5 evaluates static statistical properties of payload values within a window, including repetition, uniformity, and distributional consistency. This category captures content-level irregularities by analyzing value distributions without considering sequential ordering of frames.
\begin{tcolorbox}
\textbf{Question:} 
For any ID, is payload variability (variance across payload bytes) below the 10th percentile of the dataset baseline?

\textbf{Ground Truth:} 
\textit{True}

\end{tcolorbox}


\textbf{Category 6: Frame Format and Protocol-Level Integrity.}
Category 6 analyzes structural and protocol-compliance properties of CAN frames independent of payload semantics. It captures deviations in frame construction, field usage, and protocol-level validity that indicate abnormal or malformed communication behavior.
\begin{tcolorbox}
\textbf{Question:} 
What is the most plausible protocol-level explanation for this window, where high DLC means DLC 8 or higher and the share exceeds 50\%?

\textbf{Options:}
\begin{enumerate}[label=\Alph*.]
\item CAN arbitration effects (the high-DLC share exceeds 50\% and no non-zero Flag is present and at least one payload byte is outside its baseline range).
\item Normal change in driving conditions (the high-DLC share does not exceed 50\%).
\item Abnormal frame composition consistent with injected or altered traffic (the high-DLC share exceeds 50\% and any non-zero Flag is present).
\item Payload value scaling differences (the high-DLC share exceeds 50\% and no payload byte is outside its baseline range).
\end{enumerate}
    
\textbf{Ground Truth:} 
\textit{A}
\end{tcolorbox}


\textbf{Category 7: Expected and Safety-Critical Message Behavior.}
Category 7 evaluates whether expected or safety-critical CAN messages are absent, delayed, or exhibit abnormal communication patterns within a window. This category isolates high-impact functional behavior and distinguishes safety-relevant message dynamics from aggregate traffic characteristics. 
\begin{tcolorbox}
\textbf{Question:} 
Does any critical-control ID, defined as one of the 3 most frequent baseline IDs, appear more than 15 times or have any frame labeled as an attack with a non-zero Flag?


\textbf{Ground Truth:} 
\textit{False}

\end{tcolorbox}


\textbf{Category 8: Payload Dynamics and Baseline Plausibility.}
Category 8 analyzes temporal evolution of payload values relative to baseline behavioral models derived from reference CAN traces. It captures implausible value transitions, abnormal state dynamics, and deviations from expected temporal payload trajectories that cannot be detected through static payload statistics alone.

\begin{tcolorbox}

\textbf{Question:} 
What does the presence of rare payload transitions most strongly indicate, where rare means baseline occurrence frequency below the 5th percentile?

\textbf{Options:}
\begin{enumerate}[label=\Alph*.]
\item Definitive proof of attack (a rare transition occurs and any non-zero Flag is present).
\item Baseline modeling artifact unless repeated (a rare transition occurs only once).
\item Implausible state transitions requiring investigation (a rare transition occurs more than once).
\item Normal but infrequent behavior (no rare transition occurs).
\end{enumerate}
    
\textbf{Ground Truth:} 
\textit{D}
\end{tcolorbox}

\textbf{Category 9: Data Consistency and Logging Artifacts.}
Category 9 identifies inconsistencies introduced by data acquisition and logging processes, including duplicated frames, timestamp irregularities, and missing or corrupted entries. By explicitly modeling data quality artifacts, this category prevents collection errors from being conflated with genuine CAN communication anomalies.

\begin{tcolorbox}

\textbf{Question:} 
Do many frames share the same rounded timestamp when times are rounded to 0.01 seconds, with at least 5\% of frames in one bucket?

\textbf{Ground Truth:} 
\textit{True}

\end{tcolorbox}

\textbf{Category 10: Attack Interpretation and Decision Signals.}
Category 10 aggregates deterministically derived signals across multiple categories to support high-level interpretation of window-level behavior. Rather than reducing analysis to a single detection label, this category captures coordinated, multi-factor behavioral patterns that indicate elevated attack likelihood or complex anomalous activity.

\begin{tcolorbox}

\textbf{Question:} 
How should multiple independent anomaly signals in this window be interpreted, where signals include non-zero Flag presence, missing expected ID, rare-ID share above 30\%, high DLC share above 50\%, or a gap above 0.001 seconds?

\textbf{Options:}
\begin{enumerate}[label=\Alph*.]
\item As weak but notable concern (exactly one signal is present).
\item As coordinated anomalous behavior (at least 2 signals are present and no non-zero Flag is present).
\item As definitive proof of attack (at least 2 signals are present and a non-zero Flag is present).
\item As unrelated coincidences (zero signals are present).
\end{enumerate}

\textbf{Ground Truth:}
\textit{B}
\end{tcolorbox}

\subsection{Template-Based Question Generation}
\Design{} reformulates CAN intrusion analysis as a structured QA task over short temporal segments of CAN traffic. Each benchmark instance corresponds to a fixed-length CAN window, rendered into textual context and paired with multiple natural-language questions and automatically computed ground-truth answers. Questions are generated from predefined templates that correspond to measurable properties of the window, such as identifier frequency, traffic distribution, temporal patterns, or payload regularity. Ground-truth answers are derived deterministically from the window statistics, ensuring reproducibility and scalability. \Design{} supports True/False and multiple-choice questions. By combining diverse QA types, the benchmark evaluates not only anomaly detection but also behavioral reasoning and higher-level understanding.

\subsubsection{True/False Question Construction}
For each CAN window, we apply high-level templates that capture common in-vehicle communication patterns, including identifier activity, timing regularity, payload stability, and message dominance or rarity. Each template produces a binary statement whose truth value is computed from window statistics. 




\subsubsection{Multiple-Choice Question Construction}
Multiple-choice questions require comparative, discriminative, or interpretive reasoning over a CAN window. Each template defines a deterministically computable property, and distractors are generated using rule-based mechanisms that produce plausible but incorrect alternatives. 






\subsection{\Design{} Dataset Construction}

We construct the \Design{} benchmark using the complete Car Hacking dataset~\cite{kang2016carhacking}, which provides labeled CAN traffic spanning four attack categories: Denial-of-Service (DoS), Fuzzy, Gear, and RPM injection. Following standard preprocessing and temporal window segmentation, we transform each window into structured QA instances.

For structured selection tasks, the benchmark includes 16,564 True/False (TF) questions and 16,564 Multiple-Choice (MCQ) questions. The per-attack distribution is balanced across categories, comprising 3,664 (DoS), 3,838 (Fuzzy), 4,442 (Gear), and 4,620 (RPM) instances per question type. Each question is deterministically generated from window-level statistics and spans ten predefined reasoning categories, ensuring systematic coverage of temporal, statistical, and behavioral properties of CAN traffic.

In total, CAN-QA contains 33,128 QA instances across all attack types. Because ground-truth answers are computed directly from measurable window-level properties rather than manual annotation, dataset construction is fully reproducible. The pipeline can be seamlessly extended to new CAN traces by recomputing baseline statistics and reapplying the template rules, thereby enabling scalable benchmark expansion.

\subsection{QA Model Evaluation}

\subsubsection{Supported Model Families}

\Design{} is designed to evaluate models that process textual inputs and produce natural-language outputs. 
Accordingly, the benchmark primarily targets LLMs. Models are primarily evaluated in a zero-shot setting to assess intrinsic reasoning over CAN-bus activity represented in textual window-level form. We also explore in-context learning to obtain improved reasoning performance. 

We evaluate seven open-source LLMs in the 7B--9B parameter range: \texttt{Qwen-2-7B}, \texttt{Qwen-2.5-7B}, \texttt{Llama-3.1-8B}, \texttt{InternLM-3-8B}, \texttt{Ministral-3-8B}, \texttt{GLM-4-9B}, and \texttt{Foundation-Sec-1.1-8B}. These models are used for the primary reasoning evaluation to compare architectural differences. We focus on open-source models within a practical parameter range to enable reproducible evaluation under realistic deployment constraints, where the QA task is executed on an in-vehicle gateway or edge processor, and sensitive CAN-bus data cannot be transmitted to external cloud services.


\subsubsection{Prompting Strategy}
\label{sec:prompt}
\Design{} evaluates all models using a standardized prompting protocol. The system prompt establishes the task context by instructing the model to analyze CAN-bus communication patterns, including timestamp ordering, identifier frequency, payload stability, byte ranges, and temporal gaps between frames. The user prompt then provides a CAN window followed by a question related to the observed communication patterns. To ensure consistent responses across models, we enforce task-specific answer constraints. For True/False questions, the model must respond with either \texttt{True} or \texttt{False}. For multiple-choice questions, the model must select a single option from \texttt{A}, \texttt{B}, \texttt{C}, or \texttt{D}. We evaluate several prompting strategies.

\textbf{Zero-shot.}
In the zero-shot setting, the LLM receives only the prompt describing the CAN window and the corresponding question. This configuration measures the model's intrinsic reasoning capability over structured CAN traffic. 

\textbf{Few-shot.}
In the few-shot setting, we prepend five example question--answer pairs to the prompt to provide in-context guidance. These examples are randomly sampled from CAN windows that are disjoint from the evaluation set, ensuring that no information leakage occurs. All decoding parameters remain identical to those used in the zero-shot configuration.

\textbf{Chain-of-Thought (CoT).}
To examine if explicit reasoning improves performance, we evaluate a chain-of-thought prompting setting in which models are instructed to reason about CAN-bus behavior. The prompt encourages the model to analyze temporal ordering, identifier frequency, payload stability, and potential anomalies within the CAN window.

\textbf{Few-shot + CoT.}
We further evaluate a combined setting in which the few-shot examples include brief reasoning steps prior to the final answer. This configuration provides both in-context demonstrations and explicit reasoning traces.

An example zero-shot prompt used for the True/False task is shown below. 
Few-shot and CoT settings follow the same prompt structure, with additional in-context examples or reasoning instructions prepended to the prompt.
\begin{tcolorbox}

\textbf{System Prompt:}

You are a CAN bus intrusion-detection analyst. Study timestamp ordering, ID frequency, payload stability, byte ranges, and gaps between frames. Use those characteristics together with the statement to determine whether it is true or false. Respond with True or False only.

\medskip
\textbf{User Prompt:}

Below is a CAN bus time window. Review the sequence carefully, note anomalies or missing identifiers, and reason about the statement.

\medskip
\textbf{Statement:}

Some CAN ID transmits an identical payload across the entire window.

\medskip
\textbf{Answer:} True or False
\end{tcolorbox}

The following examples illustrate variants of the base prompt used for different prompting strategies.

\begin{tcolorbox}
\textbf{Few-shot:}

Example CAN window, statement, and correct answer pairs are inserted before the target question.

\medskip
\textbf{CoT:}

Add the instruction: \textit{``Explain your reasoning step by step before giving the final answer.''}

\medskip
\textbf{Few-shot + CoT:}

Insert several example QA pairs that include step-by-step reasoning followed by the final answer, and then present the target question using the same reasoning format.

\end{tcolorbox}




\subsubsection{Evaluation Metrics}




Model performance is evaluated using accuracy, defined as the proportion of correctly predicted answers. Ground-truth labels are generated deterministically from rule-based templates applied to statistics computed from each CAN window, so each question corresponds to a specific measurable property of the traffic.

The same metric is used for both question formats. For True/False questions, models must output either \texttt{True} or \texttt{False}. For multiple-choice questions, models must output a single option letter (\texttt{A}, \texttt{B}, \texttt{C}, or \texttt{D}). A prediction is counted as correct when the generated answer exactly matches the template-derived ground-truth label.

%% file: Results.tex


\begin{figure}[t]
    \centering    \includegraphics[width=.49\textwidth]{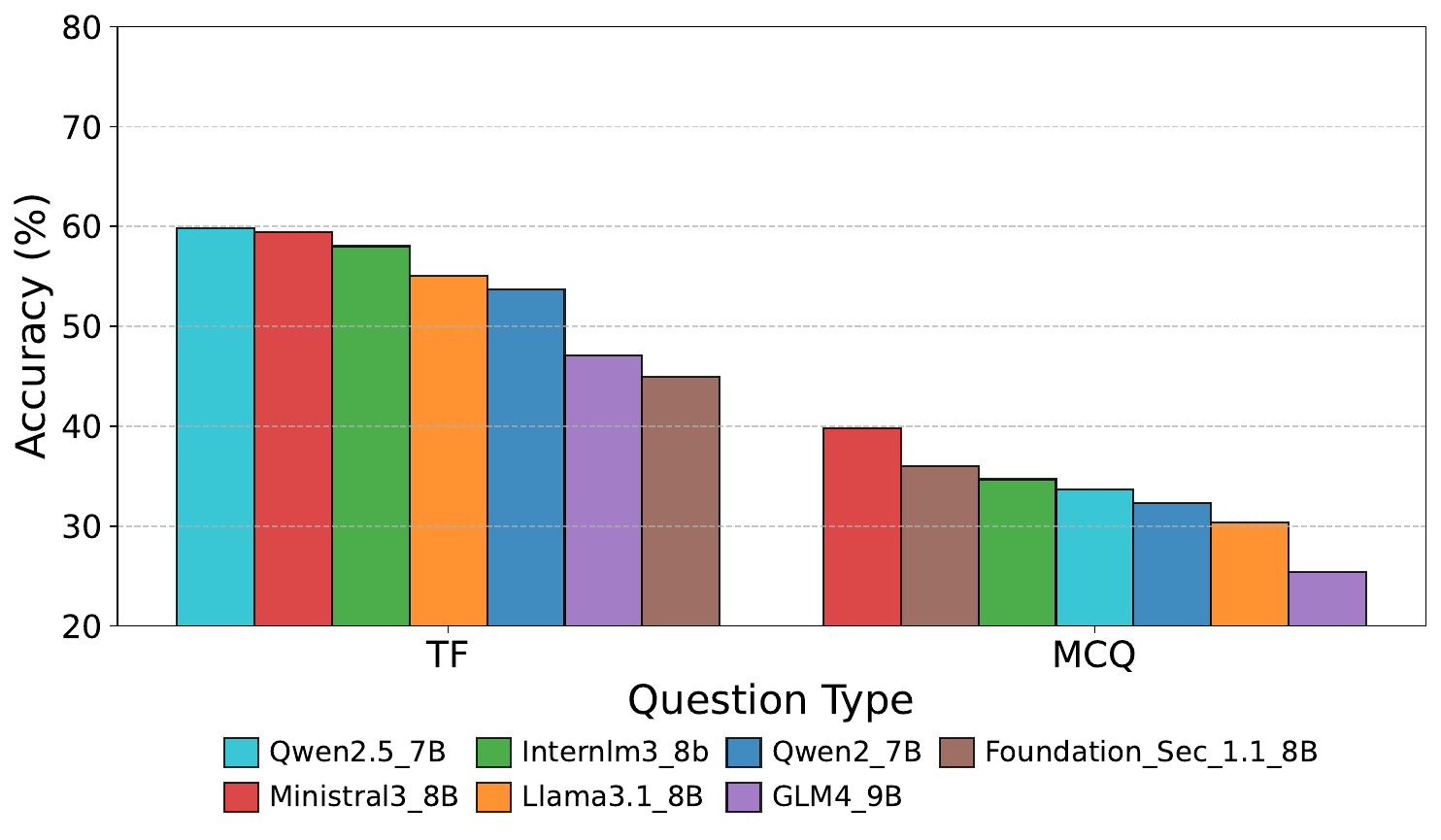}
    \caption{Zero-shot accuracy on TF and MCQ across selected models.}
    \label{fig:Accuracy_TF_MCQ}
\end{figure}

\begin{figure}[t]
    \centering
    \includegraphics[width=0.98\linewidth]{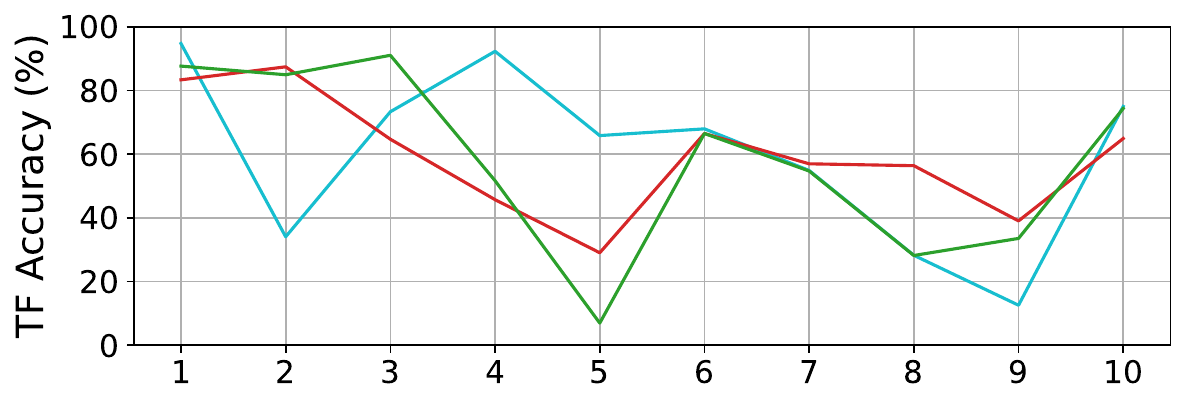}
    \vspace{0.3cm}
    \includegraphics[width=0.98\linewidth]{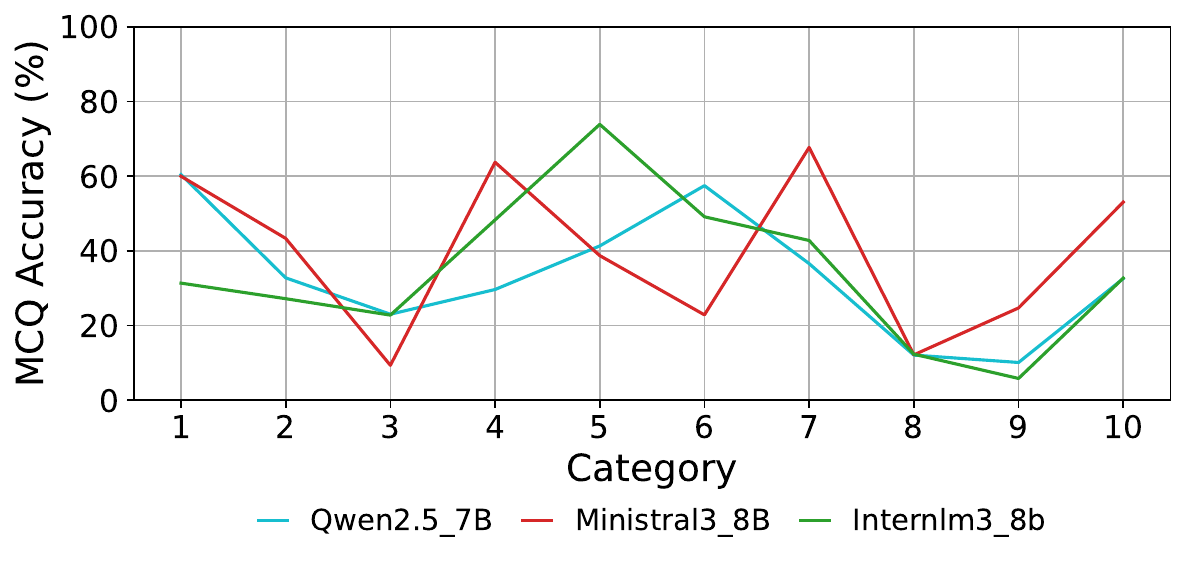}
    \caption{Per-category accuracy across TF (top) and MCQ (bottom) questions.}
    \label{fig:per_category_line}
\end{figure}




    
\subsection{Zero-Shot Performance}
Figure~\ref{fig:Accuracy_TF_MCQ} presents the zero-shot prediction accuracy of the selected LLMs on the TF and MCQ tasks. Across models, TF accuracy ranges from 47\% to 59\%, while MCQ accuracy ranges from 25\% to 40\%. Among the evaluated models, \texttt{Qwen-2.5-7B} achieves the highest accuracy on the TF task, whereas \texttt{Ministral-3-8B} performs best on MCQ. In contrast, \texttt{GLM-4-9B} records the lowest accuracy on the MCQ task, while \texttt{Foundation-Sec-1.1-8B} performs worst on TF. This result is notable given that \texttt{Foundation-Sec-1.1-8B} is specifically designed for cybersecurity applications. Despite these differences, the relative ranking of models remains largely consistent across TF and MCQ, suggesting that their reasoning capability is stable across different answer formats.

Notably, performance drops by roughly 20 percentage points when moving from TF to MCQ. This gap highlights the increased reasoning complexity of MCQ tasks. In TF questions, the model only needs to verify the correctness of a single statement. In contrast, MCQ requires selecting the most precise explanation among multiple plausible alternatives. This additional decision step increases the difficulty of the task, particularly when answer options involve subtle quantitative or structural differences. To further explore the results, we analyze performance at the category level.

\subsection{Category-Level Performance}
To examine how reasoning difficulty varies across \Design{} tasks, we analyze model accuracy across the ten question categories shown in Figure~\ref{fig:per_category_line}, where the top panel represents TF performance and the bottom panel shows MCQ accuracy. For clarity, we focus on three representative models: \texttt{Qwen-2.5-7B}, \texttt{Ministral-3-8B}, and \texttt{InternLM-3-8B}.

In the TF setting, models perform strongly on categories involving explicit statistical structure. For example, Anomalous Behavior Identification and Localization (Category 1) achieves consistently high accuracy across all three models: 94.8\% for \texttt{Qwen-2.5-7B}, 83.4\% for \texttt{Ministral-3-8B}, and 87.7\% for \texttt{InternLM-3-8B}. Strong performance is also observed in CAN Message Identity and Role Structure (Category 2) and Traffic Volume and Distribution Characteristics (Category 3). In these categories, \texttt{Ministral-3-8B} and \texttt{InternLM-3-8B} reach approximately 85–91\% accuracy, suggesting that tasks involving deterministic counting or direct threshold verification can be reliably solved using explicit statistical cues within the CAN traffic window. In contrast, categories involving payload behavior and baseline comparison are substantially more challenging. For example, Payload Value Patterns and Regularities (Category 5) exhibits strong variation across models, ranging from 6.9\% for \texttt{InternLM-3-8B} to 65.9\% for \texttt{Qwen-2.5-7B}. Similarly, Payload Dynamics and Baseline Plausibility (Category 8) remains relatively low across models in the TF setting, with accuracies of 28.3\%, 56.4\%, and 28.2\%, respectively. These tasks require reasoning relative to baseline traffic behavior rather than relying on absolute statistics. For instance, some questions ask whether payload values fall within percentile ranges defined by baseline traffic distributions, increasing interpretive complexity.

The difficulty becomes more pronounced in the MCQ task. While some categories such as Temporal Behavior of CAN Communication (Category 4) and Expected and Safety-Critical Message Behavior (Category 7) achieve moderate performance, reaching 63.7\% and 67.7\% respectively for \texttt{Ministral-3-8B}, several categories degrade sharply. In particular, Payload Dynamics and Baseline Plausibility (Category 8) becomes extremely difficult across all models, with accuracy around 12\%. 
This category requires models to reason about whether observed payload values are consistent with normal traffic baselines or represent anomalous deviations. Such reasoning often involves implicit comparisons with expected statistical ranges rather than direct threshold checks, making it difficult for models to determine which explanation best fits the observed behavior.
Data Consistency and Logging Artifacts (Category 9) also remains challenging, where performance ranges from 5.0\% for \texttt{InternLM-3-8B} to 24.7\% for \texttt{Ministral-3-8B}.
These questions typically involve identifying inconsistencies in message sequences or artifacts introduced by logging processes. Solving these problems requires models to jointly consider multiple constraints such as timing continuity, identifier repetition, and payload stability. Interestingly, some categories exhibit different difficulty patterns across TF and MCQ formats. For example, Payload Value Patterns and Regularities (Category 5) shows Low TF accuracy for certain models (e.g., 6.9\% for \texttt{InternLM-3-8B}) while achieving much higher MCQ performance (up to 73.9\%). This suggests that models may struggle to verify precise statistical conditions in declarative statements but perform better when selecting among candidate explanations that already encode plausible reasoning structures.

In contrast, Payload Dynamics and Baseline Plausibility (Category 8) remains consistently difficult across both formats. MCQ accuracy for all three models stays close to 12\%, indicating that these questions require reasoning capabilities that current models handle poorly. Such questions often involve comparisons against baseline traffic distributions and validation of multiple conditions simultaneously, which increases reasoning complexity beyond simple statistical recognition.

One explanation for these differences is the structural distinction between TF and MCQ reasoning. In TF questions, models only need to verify whether a single condition holds. In contrast, MCQ questions require selecting the most precise explanation from multiple alternatives, many of which may appear partially consistent with the observed statistics. Consequently, categories involving payload dynamics or multi-condition reasoning become substantially more challenging in the MCQ format. This is particularly evident in Temporal Behavior of CAN Communication (Category 4), Expected and Safety-Critical Message Behavior (Category 7), Payload Dynamics and Baseline Plausibility (Category 8), and Data Consistency and Logging Artifacts (Category 9). To better understand the specific mistakes models make in these challenging categories, we next perform a detailed error analysis.

\subsection{Error Analysis}
Analysis of model errors reveals recurring weaknesses in tasks that require exact numerical verification and interval-based reasoning. One common issue arises in threshold-based TF statements. Consider the example:

\begin{tcolorbox}
\textit{TF Question:}

\textit{``The number of distinct CAN IDs exceeds 30.''}
\end{tcolorbox}

The correct answer is \texttt{False}, but the model predicts \texttt{True}. This error does not indicate a complete misunderstanding of the traffic window; rather, it reflects reliance on a coarse statistical cue (e.g., the presence of many CAN IDs) without verifying whether the exact threshold is exceeded. In other words, the model is sensitive to general patterns but struggles with exact numerical verification.

A similar issue appears in MCQ questions involving adjacent quantitative intervals. For example:

\begin{tcolorbox}
\textit{MCQ Question:}

\textit{``Which traffic pattern best describes how frames are distributed across CAN IDs?''}

\textit{\textbf{Options:}}
\begin{enumerate}[label=\Alph*.]
\item \textit{Highly variable with no clear pattern (the single-appearance share exceeds 30\% and the max ID share is at most 50\%).}
\item \textit{Strongly dominated by a single ID (the max ID share exceeds 50\%).}
\item \textit{Evenly distributed across many IDs (the max ID share is at most 20\%).}
\item \textit{Moderately skewed toward a few IDs (the max ID share is above 20\% and at most 50\%).}
\end{enumerate}

\end{tcolorbox}



The correct answer is Option C (max ID share $\leq$ 20\%), while the model selects Option D (20\% $<$ max ID share $\leq$ 50\%). This error suggests that the model identifies the relevant feature dimension (ID concentration) but fails to place the observation in the correct interval. Such failures are likely because these tasks require exact boundary discrimination, whereas LLMs are often better at recognizing approximate tendencies than enforcing strict quantitative conditions. A practical way to reduce these errors would be to combine language-based reasoning with structured post-checking, such as explicit threshold validation or rule-based constraint verification after the model produces an initial interpretation.

\subsection{In-context Learning Performance}


\begin{figure}[t]
    \centering
    \includegraphics[width=.49\textwidth]{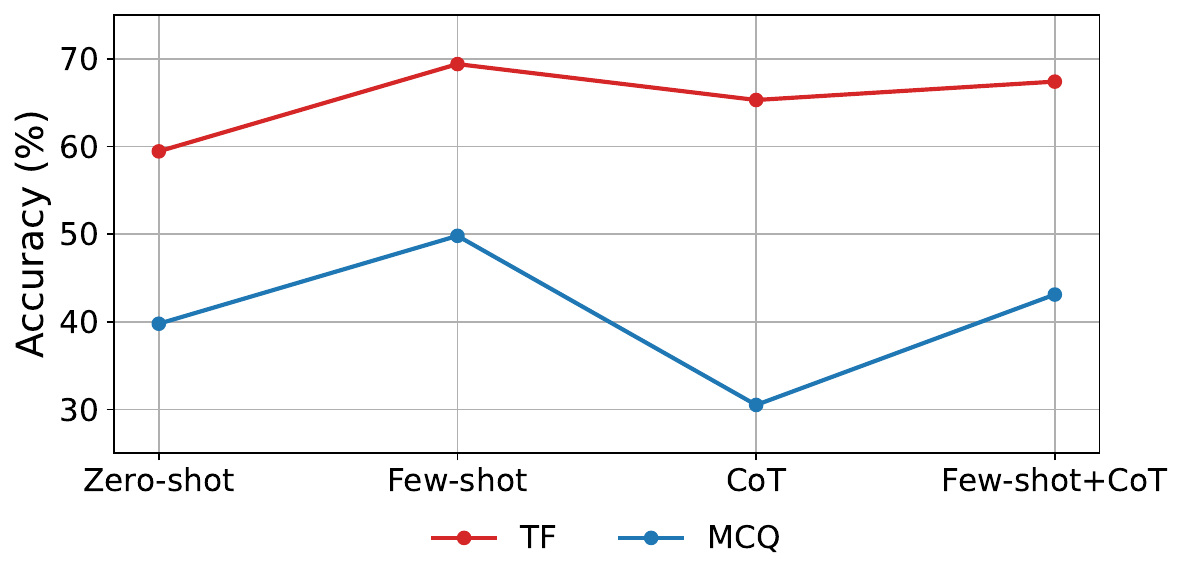}
    \caption{Effect of prompting strategies on structured CAN-QA reasoning using \texttt{Ministral-3-8B}. Performance is evaluated on True/False (TF) and Multiple-Choice (MCQ) questions under four configurations: zero-shot, few-shot, chain-of-thought (CoT), and few-shot + CoT demonstrations.}
    \label{fig:ICL_prompting}
\end{figure}




Figure~\ref{fig:ICL_prompting} compares the effect of different prompting strategies (introduced in Section~\ref{sec:prompt}) on CAN-QA reasoning for \texttt{Ministral-3-8B}, including zero-shot, few-shot, CoT, and few-shot + CoT. Few-shot prompting provides the most consistent improvement, increasing TF accuracy from roughly 59\% in the zero-shot setting to about 69\%, while also boosting MCQ performance by approximately 10 percentage points. These results suggest that providing representative examples helps the model better capture the structured reasoning patterns required for analyzing CAN traffic windows. A promising future direction is to explore adaptive few-shot prompting strategies that select the most informative examples tailored to each input window, as well as systematically investigate the effect of the number of examples on model performance.

In contrast, zero-shot CoT prompting produces mixed results. Although TF accuracy increases moderately to 65.3\%, MCQ performance drops noticeably, suggesting that generic step-by-step reasoning does not consistently align with the rule-oriented structure of CAN-QA tasks. Combining demonstrations with CoT improves over zero-shot prompting but still remains slightly weaker than few-shot alone. Overall, these results suggest that demonstrations of reasoning patterns are more effective than generic reasoning prompts. CAN-QA questions often involve structured checks such as counting events, verifying temporal relationships, and applying conditional rules, and few-shot examples provide clearer guidance for these operations than free-form chain-of-thought reasoning.

\subsection{Effect of Model Size and Generation}

\begin{figure}[t]
    \centering    \includegraphics[width=.49\textwidth]{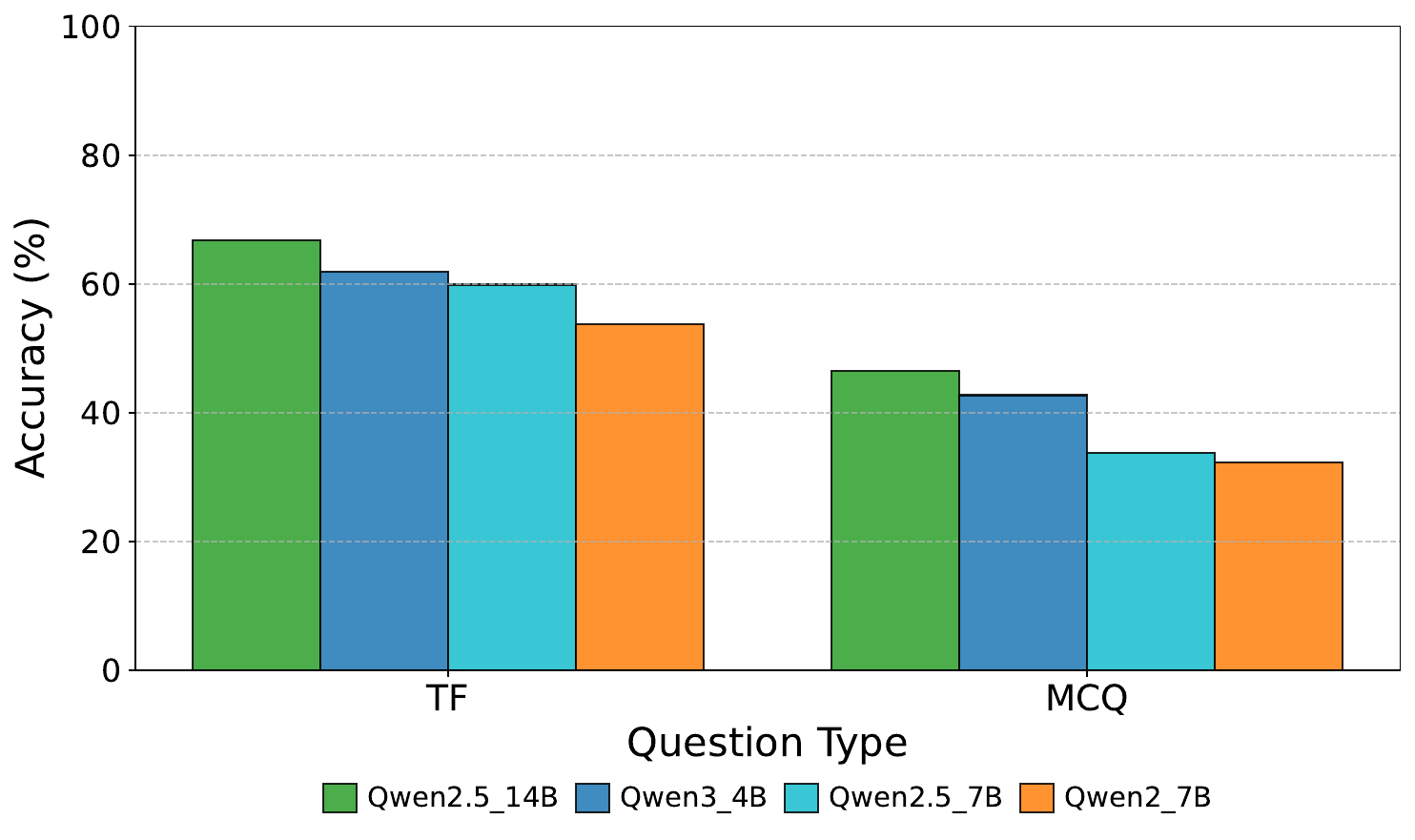}
    \caption{Structured reasoning accuracy across \texttt{Qwen} model scales (4B, 7B, and 14B). Results are reported for TF and MCQ tasks.}
    \label{fig:Model_size_Figure}
\end{figure}

To study the effects of both model size and generational improvements, we evaluate four \texttt{Qwen} variants (\texttt{Qwen-3-4B}, \texttt{Qwen-2-7B}, \texttt{Qwen-2.5-7B}, and \texttt{Qwen-2.5-14B}) under identical prompting and decoding settings, enabling controlled comparisons across parameter scaling and model iterations.

Figure~\ref{fig:Model_size_Figure} presents structured reasoning accuracy across these \texttt{Qwen} model scales. On the TF task, accuracy increases from 61.9\% (4B) to 66.7\% (14B), while MCQ performance improves from 42.7\% (4B) to 46.5\% (14B). These results suggest that larger models generally provide stronger structured reasoning capacity. However, performance does not increase monotonically across all scales: the 7B variant underperforms both the 4B and 14B models on MCQ. This indicates that reasoning performance is influenced not only by parameter count but also by architectural refinements and training improvements across model generations (\texttt{Qwen-2} vs. \texttt{Qwen-2.5} vs. \texttt{Qwen-3}). Variations in instruction tuning and alignment may also contribute to these differences.

Overall, both model scaling and iterative architectural updates play important roles in structured CAN reasoning. Larger models show measurable gains, but generational improvements can be equally significant. A promising future direction is to evaluate even larger models or ensembles of different-scale variants to investigate if combining complementary reasoning patterns can further improve structured CAN-QA performance.





\subsection{High-Level Insights}
\Design{} highlights a gap between surface-level recognition and deeper structured reasoning. Models perform reasonably well on tasks based on explicit counting or thresholds such as anomaly detection. By contrast, tasks requiring interpretation or temporal reasoning remain challenging. While models may detect unusual patterns, they often fail to produce correct interpretations, particularly in MCQ settings. Multi-condition reasoning is also limited: when aggregation of timing and payload signals is required, models respond to individual cues rather than evaluating the full set of conditions.

Overall, these findings suggest that current language models can assist analysts in summarizing observable CAN traffic patterns but are not yet reliable for safety-critical intrusion detection or full content understanding. The CAN-QA benchmark emphasizes the importance of evaluating reasoning capabilities beyond anomaly detection accuracy and highlights opportunities for future work to integrate structured time-series analysis with language-based reasoning frameworks. Such integration could enable more interpretable and robust AI-driven analysis of vehicle communication systems.






%% file: Conclusion.tex
We introduce \Design{}, the first question-answering benchmark for evaluating large language models on automotive CAN-bus traffic. By reframing intrusion analysis as a structured QA task, \Design{} goes beyond binary anomaly detection to assess a model's ability to reason over CAN behavior. The framework generates reproducible questions and ground-truth answers through rule-based templates. Our zero-shot evaluation of state-of-the-art LLMs reveals that, while models capture superficial traffic patterns, they struggle with temporal reasoning, multi-condition inference, and higher-level behavioral interpretation. Although the current framework uses fixed-length windows, deterministic templates, and a limited set of question types, it establishes a foundation for future extensions incorporating variable-length reasoning, richer question formats, and broader coverage of novel attack behaviors. Overall, \Design{} introduces a reasoning-driven evaluation paradigm that complements traditional IDS benchmarks, emphasizing structured time-series analysis, interpretable reasoning, and robust language-based modeling.